\documentclass[aps,pra,showpacs,superscriptaddress,preprint]{revtex4}
\usepackage{amssymb}
\usepackage{amsmath}
\usepackage{graphicx}

\catcode`ð=\active
 \defð{\u{g}}
 \catcode`Ð=\active
 \defÐ{\u{G}}
 \catcode`Ý=\active
\defÝ{\. I}
 \catcode`ö=\active
\defö{\"{o}}
 \catcode`Ö=\active
 \defÖ{\"O}
 \catcode`ü=\active
 \defü{\"{u}}
 \catcode`Ü=\active
 \defÜ{\"{U}}
 \catcode`Þ=\active
\defÞ{\c{S}}
 \catcode`þ=\active
 \defþ{\c{s}}
 \catcode`ý=\active
 \defý{{\i}}
 \catcode`ç=\active
\defç{\d{c}}
 \catcode`Ç=\active
\defÇ{\d{C}}

\input{tcilatex}

\begin{document}

\title{Approximate analytic solutions of the diatomic molecules in the Schr%
\"{o}dinger equation with hyperbolical potentials}
\author{Sameer M. Ikhdair}
\email[E-mail: ]{sikhdair@neu.edu.tr}
\affiliation{Department of Electrical and Electronic Engineering, Near East University,
Nicosia, Northern Cyprus, Turkey}
\author{Ramazan Sever}
\email[E-mail: ]{sever@metu.edu.tr}
\affiliation{Department of Physics, Middle East Technical University, 06531, Ankara,Turkey}
\date{%
%TCIMACRO{\TeXButton{today}{\today}}%
%BeginExpansion
\today%
%EndExpansion
}

\begin{abstract}
The Schr\"{o}dinger equation for the rotational-vibrational (ro-vibrational)
motion of a diatomic molecule with empirical potential functions is solved
approximately by means of the Nikiforov-Uvarov method. The approximate
ro-vibratinal energy spectra and the corresponding normalized total
wavefunctions are calculated in closed form and expressed in terms of the
hypergeometric functions or Jacobi polynomials $P_{n}^{\left( \mu ,\nu
\right) }(x),$ where $\mu >-1,$ $\nu >-1$ and $x\in \left[ -1,+1\right] .$
The s-waves analytic solution is obtained. The numerical energy eigenvalues
for selected $H_{2}$ and $Ar_{2}$ molecules are also calculated and compared
with the previous models and experiments.

\end{abstract}

\maketitle

\newpage

\section{Introduction}

The problems connected with the molecular sturucture provide interesting and
instructive applications of quantum mechanics, since molecules are
considerably more complex in structure than atoms. Two distict problems
arise in connection with molecular structure. The first is to obtain the
electronic wave functions and potential energy functions of the nuclear
coordinates. This problem can be solved analytically only in the simplest
cases. The second is to obtain the solution of the nuclear motion equation.
In solving the second problem, the construction of a suitable potential
function of a diatomic molecule is very important. It has been found that
the potential-energy function for the lowest electronic states of actual
diatomic molecules can be expressed by the Morse potential [1]:%
\begin{equation}
V_{M}(r)=D\left[ 1-\exp \left[ -\alpha (r-r_{e})\right] \right] ^{2},
\end{equation}%
which has three adjustable positive parameters $\alpha ,$ $D$ and $r_{e}.$
At $r=r_{e},$ it has a minimum value at zero and approaches $D$
exponentially for large $r.$ If $\frac{1}{\alpha }$ is somewhat smaller than 
$r_{e},$ it becomes large (but not infinie) as $r\rightarrow 0.$ This
potential has been the subject of many studies since 1929 [1]. Also, it is
an important potential in the field of molecular physics describing the
interaction between two atoms and has attracted a great interest for some
decades [1,2].

During the past years progress has been made in the field of diatomic
molecules and extensive use of the potential functions have been introduced
[3,4]. At present the Morse potential is still one of the potential
functions used most in molecular physics and quantum chemistry [5]. However,
the Morse potential has few asymptotic inaccuracies in the regions of small
and large $r.$ Obviously, unlike the true Coulombic interaction among two
atoms which approaches infinity, the potential is finite when the distance
of two atoms approaches zero. Another inaccuracy is the replacement of the
Van der Waals term by an exponential which makes the value of the Morse
potential smaller than that from the experiments in the region of large $r.$
To avoid these inaccuracies, many works have been carried out in that
direction to improve Morse potential [6].

In 1986, Schi\"{o}berg [7] suggested hyperbolical (empirical) potential
functions of the form: 
\begin{equation}
V_{\pm }(r)=D\left[ \delta -\sigma \left[ \coth (\alpha r)\right] ^{\pm }%
\right] ^{2},
\end{equation}%
where $D,$ $\alpha ,$ $\delta ,$ and $\sigma $ are four adjustable positive
parameter with $D=D_{e}/(\delta -\sigma )^{2}$ ($D_{e}$ is the spectroscopic
dissociation energy). It is similar to Morse potential having the minimum
value $0$ at the point%
\begin{equation}
r=r_{e}=\frac{1}{\alpha }\arctan h\left( \frac{\sigma }{\delta }\right)
^{\pm },
\end{equation}%
and approaches $D$ exponentially for large $r.$ The functional forms (2) are
significant in the limits of $0\leq \sigma <\delta $ for $V_{+}(r)$ and $%
\sigma <\delta \leq 0$ for $V_{-}(r)$. $V_{+}(r)$ becomes more and more
asymmetric if one lowers the potential constants $\delta $ and $\sigma $ but 
$V_{-}(r)$ becomes more and more symmetric with increasing $\delta $ and $%
\sigma $. One can change the potential constants of $V_{\pm }(r)$ in such a
manner that the energy eigenvalues change into all known energy values which
one obtains from rigorous quantum mechanical or semiclassical solutions.
These are the energy values of the Morse, the Kratzer, the Coulomb, the
harmonic oscillator and other potential functions. The Morse function
represents the connection point between $V_{+}(r)$ and $V_{-}(r)$ [7].
Unlike the Morse potential (1), the empirical potential (EP) function $%
V_{+}(r)$ approaches infinity at the point $r=0.$ In the region of large $r,$
it is closer to the experimental Rydberg-Klein-Rees (RKR) curve than the
Morse potential for some diatomic molecules. This potential is not only a
better description for the potential energy of a pure molecular vibrational
(= radial) energies but it represents also perfectly intermolecular
interactions and includes with the Kratzer energy eigenvalues the Rydberg
terms of an electron in an atom. Besides, the semiclassically calculated
term values for 'vibrations' in a plane there are also the
quantum-mechanical calculated term values which include these radial terms
and zero-point energies of a 'rotation' in space.

Recently, Lu [8] solved approximately the Schr\"{o}dinger equation of
diatomic molecules with the EP functions using the hypergeometric series
method. Furthermore, rigoorous solutions of the Schr\"{o}dinger eqauation
are also obtained with a similar method for zero total angular momentum.
Since there are no exact analytic solutions for the EP functions $(l\neq 0),$
some approximation [8] was used to obtain the solutions. This approximation
was employed in solving the rotating Morse potential for any $l$-states
[9,10]. The ro-vibratinal energy eigenvalues of the EP functions were
determined with a semiclassical (SC) procedure (the Bohr-Sommerfeld
quantization condition) and a quantum-mechanical (QM) method (the Schr\"{o}%
dinger equation) [8].

The NU method [11] and other methods have also been used to solve the
non-relativistic and relativistic wave equations [12-14]. The purpose of
this work is to solve the radial Scr\"{o}dinger equation\ with any orbital
angular quantum number $l$ for EP functions $V_{\pm }(r)$ using a
generalized model of the NU method derived for the exponential-type
potentials like EP. In addition, we apply the analytic solution to obtain
the ro-vibratinal energy states for selected $H_{2}$ and $Ar_{2}$ diatomic
molecules using the relevant potential parameters and spectroscpic constants
given in Ref. [7].

The present work is organized as follows. In section 2, we present a
parametric generalization of the NU method holds for any exponential-type
potential. In section 3, we obtain the analytic NU bound state solution of
the Schr\"{o}dinger equation with the EP functions for any $l$-states. In
Section 4, we calculate the ro-vibratinal energy states for selected $H_{2}$
and $Ar_{2}$ diatomic molecules. Section 5 contains the relevant conclusions.

\section{NU Method}

The Nikiforov-Uvarov (NU) method is briefly outlined here and the details
can be found in [11]. It is proposed to solve the second-order linear
differential equation by reducing it to a generalized equation of
hypergeometric-type of the following form: 
\begin{equation}
R^{\prime \prime }(z)+\left( \frac{\widetilde{\tau }(z)}{\sigma (z)}\right)
R^{\prime }(z)+\left( \frac{\widetilde{\sigma }(z)}{\sigma ^{2}(z)}\right)
R(z)=0,
\end{equation}%
where the prime denotes the differentiation with respect to $z,$ $\sigma (z)$
and $\widetilde{\sigma }(z)$ are polynomials, at most second-degree, and $%
\widetilde{\tau }(s)$ is a first-degree polynomial. In order to find a
particular solution of Eq. (4), we decompose the wavefunction $R(z)$ as
follows:%
\begin{equation}
R(z)=\phi (z)y_{n}(z),
\end{equation}%
which reduces Eq. (4) to a hypergeometric type equation 
\begin{equation}
\sigma (z)y_{n}^{\prime \prime }(z)+\tau (z)y_{n}^{\prime }(z)+\lambda
y_{n}(z)=0,
\end{equation}%
where $\phi (z)$ is defined as a logarithmic derivative%
\begin{equation}
\phi ^{\prime }(z)/\phi (z)=\pi (z)/\sigma (z),
\end{equation}%
and the other part $y_{n}(z)$ is the hypergeometric-type function whose
polynomial solution satisfies the Rodrigues relation:%
\begin{equation}
y_{n}(z)=\frac{A_{n}}{\rho (z)}\frac{d^{n}}{dz^{n}}\left[ \sigma ^{n}(z)\rho
(z)\right] ,
\end{equation}%
where $A_{n}$ is a normalizing factor and $\rho (z)$ is the weight function
satisfying the condition 
\begin{equation}
\left( \sigma (z)\rho (z)\right) ^{\prime }=\tau (z)\rho (z).
\end{equation}
The function $\pi (z)$ and the eigenvalue $\lambda $ required in this method
are defined as%
\begin{equation}
\pi (z)=\frac{1}{2}\left[ \sigma ^{\prime }(z)-\widetilde{\tau }(z)\right]
\pm \sqrt{\frac{1}{4}\left[ \sigma ^{\prime }(z)-\widetilde{\tau }(z)\right]
^{2}-\widetilde{\sigma }(z)+k\sigma (z)}.
\end{equation}
and 
\begin{equation}
\lambda =k+\pi ^{\prime }(z).
\end{equation}%
Hence, the determination of $k$ is the essential point in the calculation of 
$\pi (z),$ for which the discriminant of the square root in Eq. (10) is set
to zero. Also, the eigenvalue equation defined in Eq. (11) takes the
following new form%
\begin{equation}
\lambda =\lambda _{n}=-n\tau ^{\prime }(z)-\frac{1}{2}n\left( n-1\right)
\sigma ^{\prime \prime }(z),\ \ \ n=0,1,2,\cdots .
\end{equation}%
where%
\begin{equation}
\tau (z)=\widetilde{\tau }(z)+2\pi (z),
\end{equation}%
and it's derivative is negative (i.e., $\tau ^{\prime }(z)<0$) which is the
essential condition for any choice of proper bound-state solutions. Finally,
comparing Eqs. (11) and (12), we obtain the energy eigenalues.

In this regard, we can derive a parametric generalization of the NU method
valid for any central and non-central exponential-type potential [12]. We
begin by comparing comparing the following generalized hypergeometric-type
equation 
\begin{equation}
R^{\prime \prime }(z)+\frac{\left( c_{1}-c_{2}z\right) }{z\left(
1-c_{3}z\right) }R^{\prime }(z)+\frac{1}{\left[ z\left( 1-c_{3}z\right) %
\right] ^{2}}\left( -\xi _{1}z^{2}+\xi _{2}z-\xi _{3}\right) R(z)=0,
\end{equation}%
with Eq. (4), we obtain 
\begin{subequations}
\begin{equation}
\widetilde{\tau }(z)=c_{1}-c_{2}z,
\end{equation}%
\begin{equation}
\sigma (z)=z\left( 1-c_{3}z\right)
\end{equation}%
\begin{equation}
\widetilde{\sigma }(z)=-\xi _{1}z^{2}+\xi _{2}z-\xi _{3}.
\end{equation}%
Further, substituting Eqs. (15a)-(15c) into Eq. (10), we find 
\end{subequations}
\begin{equation}
\pi (z)=c_{4}+c_{5}z\pm \left[ \left( c_{6}-c_{3}k_{+,-}\right) z^{2}+\left(
c_{7}+k_{+,-}\right) z+c_{8}\right] ^{1/2},
\end{equation}%
where%
\begin{equation}
c_{4}=\frac{1}{2}\left( 1-c_{1}\right) ,\text{ }c_{5}=\frac{1}{2}\left(
c_{2}-2c_{3}\right) ,\text{ }c_{6}=c_{5}^{2}+\xi _{1},\text{ }%
c_{7}=2c_{4}c_{5}-\xi _{2},\text{ }c_{8}=c_{4}^{2}+\xi _{3}.
\end{equation}%
The discriminant under the square root sign in Eq. (16) must be set to zero
and the resulting equation has to be solved for $k,$ it yields%
\begin{equation}
k_{+,-}=-\left( c_{7}+2c_{3}c_{8}\right) \pm 2\sqrt{c_{8}c_{9}},
\end{equation}%
where%
\begin{equation}
c_{9}=c_{3}\left( c_{7}+c_{3}c_{8}\right) +c_{6}.
\end{equation}%
Substituting Eq. (18) into Eq. (16) and then solving, we obtain the physical
choice of parameters: 
\begin{equation}
\pi (z)=c_{4}+c_{5}z-\left[ \left( \sqrt{c_{9}}+c_{3}\sqrt{c_{8}}\right) z-%
\sqrt{c_{8}}\right] ,
\end{equation}%
\begin{equation}
k_{-}=-\left( c_{7}+2c_{3}c_{8}\right) -2\sqrt{c_{8}c_{9}}.
\end{equation}%
In addition, Eqs. (13), (15) and (20) provide the parameter 
\begin{equation}
\tau (z)=1-\left( c_{2}-2c_{5}\right) z-2\left[ \left( \sqrt{c_{9}}+c_{3}%
\sqrt{c_{8}}\right) z-\sqrt{c_{8}}\right] ,
\end{equation}%
whose derivative must be negative:%
\begin{equation}
\tau ^{\prime }(z)=-2c_{3}-2\left( \sqrt{c_{9}}+c_{3}\sqrt{c_{8}}\right) <0,
\end{equation}%
in accordance with essential requirement of the method [11]. Solving Eqs.
(11) and (12), we obtain the parametric energy equation:%
\begin{equation}
\left( c_{2}-c_{3}\right) n+c_{3}n^{2}-\left( 2n+1\right) c_{5}+\left(
2n+1\right) \left( \sqrt{c_{9}}+c_{3}\sqrt{c_{8}}\right) +c_{7}+2c_{3}c_{8}+2%
\sqrt{c_{8}c_{9}}=0,
\end{equation}%
for the exponential-type potential under study. Let us now turn to the
calculations of the wavefunctions. The weight function $\rho (z)$ can be
calculated by means of Eq. (9) as%
\begin{equation}
\rho (z)=z^{c_{10}}(1-c_{3}z)^{c_{11}},
\end{equation}%
and consequently the first part of the wavefunctions throughout the
Rodrigues relation (8):%
\begin{equation}
y_{n}(z)=P_{n}^{\left( c_{10},c_{11}\right) }(1-2c_{3}z),\text{ }c_{10}>-1,%
\text{ }c_{11}>-1,\text{ }z\in \left[ 0,1/c_{3}\right] ,
\end{equation}%
where 
\begin{equation}
c_{10}=c_{1}+2c_{4}+2\sqrt{c_{8}}-1>-1,\text{ }c_{11}=1-c_{1}-2c_{4}+\frac{2%
}{c_{3}}\sqrt{c_{9}}>-1,
\end{equation}%
and $P_{n}^{\left( a,b\right) }(1-2c_{3}z)$ are the Jacobi polynomials.
Also, the second part of the wavefunctions can be found from (7) as%
\begin{equation}
\phi (z)=z^{c_{12}}(1-c_{3}z)^{c_{13}},\text{ }c_{12}>0,\text{ }c_{13}>0,
\end{equation}%
where 
\begin{equation}
c_{12}=c_{4}+\sqrt{c_{8}}>0,\text{ }c_{13}=-c_{4}+\frac{1}{c_{3}}\left( 
\sqrt{c_{9}}-c_{5}\right) >0.
\end{equation}%
Hence, the general wavefunction (5) has the general form%
\begin{equation}
R(z)=\mathcal{N}_{n}z^{c_{12}}(1-c_{3}z)^{c_{13}}P_{n}^{\left(
c_{10},c_{11}\right) }(1-2c_{3}z),
\end{equation}%
where $\mathcal{N}_{n}$ is a normalization constant.

\section{Bound States of the Hyperbolical Potentials}

The Schr\"{o}dinger equation for diatomic molecules with hyperbolical
potential functions takes the form:%
\begin{equation}
\left\{ \mathbf{\nabla }^{2}+\frac{2\mu }{\hbar ^{2}}\left[ E_{nl}-V_{\pm
}(r)\right] \right\} \psi _{nlm}(r,\theta ,\varphi )=0,
\end{equation}%
where $\mu =\frac{m_{1}m_{2}}{m_{1}+m_{2}}.$ The wavefunctions in the above
equation could be separated to the following form [15-18] 
\begin{equation}
\psi _{nlm}(r,\theta ,\varphi )=\frac{1}{r}R_{nl}(r)Y_{lm}(\theta ,\varphi ),
\end{equation}%
where $Y_{lm}(\theta ,\varphi )$ is a spherical harmonic with angular
momentum quantum numbers $l$ and $m.$ The substitution of Eq. (32) into Eq.
(31) gives the following radial reduced wave function $R_{nl}(r)$ satisfying%
\begin{equation}
\left\{ \frac{d^{2}}{dr^{2}}-\frac{l(l+1)}{r^{2}}+\frac{2\mu }{\hbar ^{2}}%
\left[ E_{nl}-V_{\pm }(r)\right] \right\} R_{nl}(r)=0,
\end{equation}%
where $\frac{l(l+1)}{r^{2}}$ is the centrifugal potential with the boundary
condition that $R_{l}(r)$ vanishes near the points $r=0$ and $r\rightarrow
\infty .$ Furthermore, if $l$ is not too large, the case of the vibrations
of small amplitude about the minimum, we can then use the approximate
expansion of the centrifugal potential near the minimum point $r=r_{e}$ as
[19]:%
\begin{equation}
\frac{l(l+1)}{r^{2}}\approx \frac{l(l+1)}{r_{e}^{2}}\left\{ A_{0}+A_{1}\frac{%
\pm \exp (-2\alpha r)}{1\mp \exp (-2\alpha r)}+A_{2}\left[ \frac{\pm \exp
(-2\alpha r)}{1\mp \exp (-2\alpha r)}\right] ^{2}\right\} ,
\end{equation}%
where 
\begin{subequations}
\begin{equation}
A_{0}=1-\left[ \frac{1\mp \exp (-2\alpha r_{e})}{2\alpha r_{e}}\right] ^{2}%
\left[ \frac{8\alpha r_{e}}{1\mp \exp (-2\alpha r_{e})}-\left( 3+2\alpha
r_{e}\right) \right] ,
\end{equation}%
\begin{equation}
A_{1}=\pm 2\left[ \exp (2\alpha r_{e})\mp 1\right] \left\{ 3\left[ \frac{%
1\mp \exp (-2\alpha r_{e})}{2\alpha r_{e}}\right] -\left( 3+2\alpha
r_{e}\right) \left[ \frac{1\mp \exp (-2\alpha r_{e})}{2\alpha r_{e}}\right]
^{2}\right\} ,
\end{equation}%
\begin{equation}
A_{2}=\left[ \exp (2\alpha r_{e})\mp 1\right] ^{2}\left[ \frac{1\mp \exp
(-2\alpha r_{e})}{2\alpha r_{e}}\right] ^{2}\left[ 3+2\alpha r_{e}-\frac{%
4\alpha r_{e}}{1\mp \exp (-2\alpha r_{e})}\right] ,
\end{equation}%
and higher order terms are neglected. In fact, Eq. (34) is the approximate
expansion of the centrifugal potential $l(l+1)r^{-2}$ and is valid for all $%
r\approx r_{e}$, the minimum point of $V_{\pm }(r)$ since $r$ is not
singular there$.$ However, the expansion is not valid near the singularity
point $r=0.$ Overmore, it is a good approximation for small vibrations
around the equilibrium separation $r-r_{e}.$

Substituting Eqs. (2) and (34) into Eq. (33) and introducing a new variable $%
z=\pm \exp (-2\alpha r),$ where $z\in $ [$\pm 1,0]$ for $V_{\pm }(r),$ we
obtain the hypergeometric-type wave equation: 
\end{subequations}
\begin{equation*}
R_{nl}^{\prime \prime }(z)+\frac{\left( 1-z\right) }{z\left( 1-z\right) }%
R_{nl}^{\prime }(z)+\frac{1}{z^{2}\left( 1-z\right) ^{2}}
\end{equation*}%
\begin{equation}
\times \left[ -\left( K_{nl}^{2}+S_{l}^{2}-Q_{l}-\frac{1}{4}\right)
z^{2}+\left( 2K_{nl}^{2}-Q_{l}\right) z-K_{nl}^{2}\right] R_{nl}(z)=0,
\end{equation}%
where $R_{nl}(z)=R_{nl}(r)$ and 
\begin{subequations}
\begin{equation}
K_{nl}=\sqrt{\frac{\mu D}{2\alpha ^{2}\hbar ^{2}}\left( \delta -\sigma
\right) ^{2}+\frac{l(l+1)}{4\alpha ^{2}r_{e}^{2}}A_{0}-\frac{\mu E_{nl}}{%
2\alpha ^{2}\hbar ^{2}}},
\end{equation}%
\begin{equation}
Q_{l}=-\frac{2\mu D}{\alpha ^{2}\hbar ^{2}}\sigma \left( \delta -\sigma
\right) +\frac{l(l+1)}{4\alpha ^{2}r_{e}^{2}}A_{1},
\end{equation}%
\begin{equation}
S_{l}=\sqrt{\frac{2\mu D}{\alpha ^{2}\hbar ^{2}}\sigma ^{2}+\frac{l(l+1)}{%
4\alpha ^{2}r_{e}^{2}}A_{2}+\frac{1}{4}}.
\end{equation}%
By comparing Eq. (36) with Eq. (14), we obtain specific values for the set
of constant parameters given in Section 2: 
\end{subequations}
\begin{equation*}
c_{1}=\text{ }c_{2}=c_{3}=1,\text{ c}_{4}=0,\text{ }c_{5}=-\frac{1}{2},\text{
}c_{6}=K_{nl}^{2}+S_{l}^{2}-Q_{l},
\end{equation*}%
\begin{equation*}
c_{7}=-2K_{nl}^{2}+Q_{l},\text{ }c_{8}=K_{nl}^{2},\text{ }c_{9}=S_{l}^{2},
\end{equation*}%
\begin{equation*}
c_{10}=2K_{nl},\text{ }c_{11}=2S_{l},\text{ }c_{12}=K_{nl},\text{ }%
c_{13}=S_{l}+\frac{1}{2},
\end{equation*}%
\begin{equation}
\xi _{1}=K_{nl}^{2}+S_{l}^{2}-Q_{l}-\frac{1}{4},\text{ }\xi
_{2}=2K_{nl}^{2}-Q_{l},\text{ }\xi _{3}=K_{nl}^{2}.
\end{equation}%
By using Eqs. (20)-(22), we find the following physical values:%
\begin{equation}
\pi (z)=K_{nl}-\left( \frac{1}{2}+K_{nl}+S_{l}\right) z,
\end{equation}%
\begin{equation}
k=-Q_{l}-2K_{nl}S_{l},
\end{equation}%
\begin{equation}
\tau (z)=1+2K_{nl}-2\left( 1+K_{nl}+S_{l}\right) z,
\end{equation}%
where $\tau ^{\prime }(z)=-2\left( 1+K_{nl}+S_{l}\right) <0$ is the
essential condition for bound-state (real) solutions$.$ In addition, the
energy equation can be found via Eq. (24) as

\begin{equation}
2K_{nl}=\frac{S_{l}^{2}-Q_{l}-\frac{1}{4}-\left( S_{l}+n+\frac{1}{2}\right)
^{2}}{S_{l}+n+\frac{1}{2}}.
\end{equation}%
The energy eigenvalues are obtained as follows,%
\begin{equation*}
E_{nl}=D\left( \delta -\sigma \right) ^{2}+\frac{l(l+1)\hbar ^{2}}{2\mu
r_{e}^{2}}A_{0}
\end{equation*}%
\begin{equation*}
-\frac{\hbar ^{2}\alpha ^{2}}{2\mu }\left[ \frac{S_{l}^{2}-Q_{l}-\frac{1}{4}%
-\left( S_{l}+n+\frac{1}{2}\right) ^{2}}{S_{l}+n+\frac{1}{2}}\right] ^{2},%
\text{ }n,l=0,1,2,\cdots ,
\end{equation*}%
where $n$ and $l$ signify the usual vibrational and rotational quantum
numbers, respectively. Thus, the ro-vibrational energy spectrum takes the
following explicit form 
\begin{equation*}
E_{nl}=D_{e}+\frac{l(l+1)\hbar ^{2}}{2\mu r_{e}^{2}}A_{0}
\end{equation*}%
\begin{equation}
-\frac{\hbar ^{2}\alpha ^{2}}{2\mu }\left[ \frac{\frac{2\mu D}{\hbar
^{2}\alpha ^{2}}\sigma \delta +\frac{l(l+1)}{4\alpha ^{2}r_{e}^{2}}\left(
A_{2}-A_{1}\right) -\left( n+\frac{1}{2}+\sqrt{\frac{2\mu D}{\hbar
^{2}\alpha ^{2}}\sigma ^{2}+\frac{l(l+1)}{4\alpha ^{2}r_{e}^{2}}A_{2}+\frac{1%
}{4}}\right) ^{2}}{n+\frac{1}{2}+\sqrt{\frac{2\mu D}{\hbar ^{2}\alpha ^{2}}%
\sigma ^{2}+\frac{l(l+1)}{4\alpha ^{2}r_{e}^{2}}A_{2}+\frac{1}{4}}}\right]
^{2},
\end{equation}%
which is identical to Eq. (28) of Ref. [8]. Its important to mention that
very similar expressions to the above expression for the energy states have
been found over the past years for the hypebolical (exponential-type)
potentials with $\delta $ is being set equal to one in Eq. (2) (cf.
[19-21]). Very recently, a new improved approximation [21,22] for the
centrifugal potential term $l(l+1)/r^{2}$ was used different than the ones
used in Ref. [8] and the one which is commonly used in literature by Ref.
[23].

Let us now turn to the calculations of the corresponding wavefunctions for
the EP functions. Thus, referring to the parametric generalization of the NU
method in Section 2, the weight function in Eq. (25) takes the form 
\begin{equation}
\rho (z)=z^{2K_{nl}}(1-z)^{2S_{l}},
\end{equation}%
which gives the first part of the wavefunctions in Eq. (5) as%
\begin{equation}
y_{n}(z)\rightarrow P_{n}^{(2K_{nl},2S_{l})}(1-2z).
\end{equation}%
Also, the second part of the wavefunctions (28) can be found as%
\begin{equation}
\phi (z)\rightarrow z^{K_{nl}}(1-z)^{S_{l}+\frac{1}{2}}.
\end{equation}%
Hence, the unnormalized wavefunctions (30) are being expressed in terms of
the Jacobi polynomials as%
\begin{equation}
R_{nl}(z)=\mathcal{N}_{nl}z^{K_{nl}}(1-z)^{S_{l}+\frac{1}{2}%
}P_{n}^{(2K_{nl},2S_{l})}(1-2z),
\end{equation}%
where $\mathcal{N}_{nl}$ is a normalizing factor and $%
P_{n}^{(2K_{nl},2S_{l})}(1-2z)=\frac{\left( 2K_{nl}+1\right) _{n}}{n!}%
_{2}F_{1}(-n,2K_{nl}+2S_{l}+n+1,2K_{nl}+1;z)$ with $(m)_{n}=\frac{\left(
m+n-1\right) !}{\left( m-1\right) !}$ is Pochhammer's symbol$.$

Hence, the total wavefunctions of the EP are 
\begin{equation*}
\psi _{\pm }(r,\theta ,\varphi )=\mathcal{N}_{nl}\frac{1}{r}\left[ \pm \exp
(-2\alpha r)\right] ^{K_{nl}}\left[ 1-\pm \exp (-2\alpha r)\right] ^{S_{l}+%
\frac{1}{2}}
\end{equation*}%
\begin{equation}
\times P_{n}^{(2K_{nl},2S_{l})}(1-\pm 2\exp (-2\alpha r))Y_{lm}(\theta
,\varphi ).
\end{equation}%
where the normalization constants $\mathcal{N}_{nl}$ are determined in
Appendix A$.$

Let us find the vibrational energy states for the $s$-waves $(l=0)$ from Eq.
(43) as 
\begin{equation}
E_{n}=D_{e}-\frac{\hbar ^{2}\alpha ^{2}}{2\mu }\left[ \frac{\frac{2\mu D}{%
\hbar ^{2}\alpha ^{2}}\sigma \delta -\left( \sqrt{\frac{2\mu D}{\hbar
^{2}\alpha ^{2}}\sigma ^{2}+\frac{1}{4}}+n+\frac{1}{2}\right) ^{2}}{\sqrt{%
\frac{2\mu D}{\hbar ^{2}\alpha ^{2}}\sigma ^{2}+\frac{1}{4}}+n+\frac{1}{2}}%
\right] ^{2},\text{ }n=0,1,2,\cdots ,n_{\max },
\end{equation}%
where $n_{\max }$ is the number of bound states for the whole bound spectrum
near the continuous zone. $n_{\max }$ is the largest integer which is less
than or equal to the value of $n$ that makes the right side of Eq. (49)
vanish, that is,%
\begin{equation}
n_{\max }=\frac{1}{2}\left( \sqrt{\frac{8\mu D}{\hbar ^{2}\alpha ^{2}}\sigma
\delta }-\sqrt{\frac{8\mu D}{\hbar ^{2}\alpha ^{2}}\sigma ^{2}+1}-1\right) .
\end{equation}%
Thus, $n_{\max }$ cannot be infinite ($E_{n_{\max }}=D_{e}$), which is
reflected in the above condition. Furthermore, the corresponding
wavefunctions for the $s$-waves can be easily found from Eqs. (37) and (48)
as 
\begin{equation*}
\psi _{\pm }(r,\theta ,\varphi )=\mathcal{N}_{n}\frac{1}{r}\left[ \pm \exp
(-2\alpha r)\right] ^{k_{n}}\left[ 1-\pm \exp (-2\alpha r)\right] ^{s+\frac{1%
}{2}}
\end{equation*}%
\begin{equation}
\times P_{n}^{(2k_{n},2s)}(1-\pm 2\exp (-2\alpha r))Y_{0,0}(\theta ,\varphi
),
\end{equation}%
where the normalization constants $\mathcal{N}_{n}$ are determined in
Appendix A.

\section{Applications to Diatomic Molecules}

In this section, we calculate the energy states for the two selected $H_{2}$
and $Ar_{2}$ diatomic molecules using Eqs. (35) and (43). The spectroscopic
constants of these two molecules are given in Table 1. The vibrating ground
state energy eigenvalues $E_{+}^{00}$ (in $cm^{-1}$) for the $H_{2}$
molecule in the EP functions $V_{+}(r)$ are found by means of parametric
generalization version of the NU method for the potential parameters given
in Table 2. Our numerical results obtained in the present NU model are
listed together with the analogous numerical results obtained by using SC
procedure and a QM method mentioned in Ref. [7] for various potential
parameters. Obviously, as shown in Table 2, the results obtained in the
present model are in high agreement with those obtained by QM, however, the
SC procedure is proportionally different. Therefore, the differences between
our results and SC procedure are less than $0.01$ $cm^{-1},$ i.e., they are
negligible because of these approximations: $1$ $a.m.u=931.502$ $MeV/c^{2},$ 
$1$ $cm^{-1}=1.23985\times 10^{-4}$ $eV$ and $\hbar c=1973.29$ $eV.A^{\circ
} $ [28]. The second application is applied to $Ar_{2}$ molecule. We confine
our study to calculate the ro-vibrating energy states for the $V_{+}(r)$
potential using the the following potential parameters: $\sigma =25.23,$ $%
\delta =41.75$ and $\alpha =0.6604$ $(A^{\circ })^{-1}$ [7] together with
the parameters given in Table 1. The splittings of the energy states of $s$%
-waves $E_{+}=E_{+}(n\neq 0)-E_{+}(n=0)$ obtained by the NU method and SC
procedures are presented in Table 3. The present results $\Delta E_{+}(NU)$
from NU method and $\Delta E_{+}(SC)$ obtained from the SC procedures are
also compared with four-different experimental results labeled by $\Delta
E(a),\Delta E(b),$ $\Delta E(c)$ and $\Delta E(d)$ taken from Ref. [7]$.$ It
is clear from Table 3 that our results are very close with the
expremimentally determined values as well as the SC procedure results.
Finally, the approximated ro-vibrating energy states of the $V_{+}(r)$ given
in Eq. (2) for the $Ar_{2}$ and $H_{2}$ molecules are also calculated for
the $l\neq 0$ case. Table 4 shows the energy levels for vibrational $%
(n=0,1,2,3,4,5)$ and rotational $(l=0,1,2)$ quantum numbers.

\section{Conclusions}

In the present work, we have used a parametric generalization version of the
NU method derived for any exponential-type potential to obtain the
approximate solutions of the Schr\"{o}dinger equation with any orbital
angular momentum quantum number $l$ for the hyperbolical (EP) functions.
This method is systematical and efficient in finding the ro-vibrating energy
states of a diatomic molecule and the normalized wave functions expressed in
terms of the Jacobi polynomials. Obviously, the energy eigenvalues equation
given in Eq. (43), considering the approximation in Eq. (35), is identical
to those obtained by functional analysis method [8]. The analytical result
is tested in calculating the ro-vibrating energy states of the $H_{2}$ and $%
Ar_{2}$ molecules. Comparisons with the results of previous methods SC and
QM for the ground state $(n=l=0)$ show that our calculations are in high
agreement with those experimental results for $H_{2}$ and $Ar_{2}$ for low
values of the ro-vibrating energy states, i.e., $l=0,1,2.$ These
systematical procedures could be useful for other molecular potentials in
calculating their higher or lower approximated energy states [29-31].

\acknowledgments The partial support provided by the Scientific and
Technological Research Council of Turkey (T\"{U}B\.{I}TAK) is highly
appreciated.

\appendix

\section{Normalization of the radial wave function}

In order to find the normalization constants $\mathcal{N}_{nl}$, we start by
writting the normalization condition:%
\begin{equation}
\mathcal{N}_{nl}^{-2}=\frac{1}{2\alpha }%
\int_{0}^{1}z^{2K_{nl}-1}(1-z)^{2S_{l}+1}\left[
P_{n}^{(2K_{nl},2S_{l})}(1-2z)\right] ^{2}dz.
\end{equation}%
Unfortunately, there is no formula available to calculate this key
integration. Neveretheless, we can find the explicit normalization constants 
$\mathcal{N}_{nl}.$ For this purpose, it is not difficult to obtain the
results of the above integral by using the following formulas [24-27,29-31]%
\begin{equation}
\dint\limits_{0}^{1}\left( 1-z\right) ^{\mu -1}z^{\nu -1}%
\begin{array}{c}
_{2}F_{1}%
\end{array}%
\left( \alpha ,\beta ;\gamma ;az\right) dz=\frac{\Gamma (\mu )\Gamma (\nu )}{%
\Gamma (\mu +\nu )}%
\begin{array}{c}
_{3}F_{2}%
\end{array}%
\left( \nu ,\alpha ,\beta ;\mu +\nu ;\gamma ;a\right) ,
\end{equation}%
and $%
\begin{array}{c}
_{2}F_{1}%
\end{array}%
\left( a,b;c;z\right) =\frac{\Gamma (c)}{\Gamma (a)\Gamma (b)}%
\dsum\limits_{p=0}^{\infty }\frac{\Gamma (a+p)\Gamma (b+p)}{\Gamma (c+p)}%
\frac{z^{p}}{p!}.$ Following Ref. [24-27,29-31], we calculate the
normalization constants:%
\begin{equation}
\mathcal{N}_{nl}=\left[ \frac{\Gamma (2K_{nl}+1)\Gamma (2S_{l}+2)}{2\alpha
\Gamma (n)}\dsum\limits_{m=0}^{\infty }\frac{(-1)^{m}\left(
1+n+2(K_{nl}+S_{l})\right) _{m}\Gamma (n+m)}{m!\left( m+2K_{nl}\right)
!\Gamma \left( m+2(K_{nl}+S_{l}+1)\right) }f_{nl}\right] ^{-1/2}\text{ ,}
\end{equation}%
where 
\begin{equation}
f_{nl}=%
\begin{array}{c}
_{3}F_{2}%
\end{array}%
\left(
2K_{nl}+m,-n,n+1+2(K_{nl}+S_{l});m+2(K_{nl}+S_{l}+1);1+2K_{nl};1\right) .
\end{equation}%
Furthermore, the normalization constants for the $s$-wave can be also found
as%
\begin{equation}
\mathcal{N}_{n}=\left[ \frac{\Gamma (2k_{n}+1)\Gamma (2s+2)}{2\alpha \Gamma
(n)}\dsum\limits_{m=0}^{\infty }\frac{(-1)^{m}\left( 1+n+2(k_{n}+s)\right)
_{m}\Gamma (n+m)}{m!\left( m+2k_{n}\right) !\Gamma \left(
m+2(k_{n}+s+1)\right) }g_{n}\right] ^{-1/2}\text{ ,}
\end{equation}%
where 
\begin{equation}
g_{n}=%
\begin{array}{c}
_{3}F_{2}%
\end{array}%
\left( 2k_{n}+m,-n,n+1+2(k_{n}+\widetilde{s});m+2(k_{n}+s+1);1+2k_{n};1%
\right) ,
\end{equation}%
and 
\begin{equation}
k_{n}=\sqrt{\frac{\mu D}{2\hbar ^{2}\alpha ^{2}}\left( \delta -\sigma
\right) ^{2}-\frac{\mu E_{n}}{2\hbar ^{2}\alpha ^{2}}},\text{ }s=\sqrt{\frac{%
2\mu D}{\hbar ^{2}\alpha ^{2}}\sigma ^{2}+\frac{1}{4}},\text{ }%
n=0,1,2,\cdots ,
\end{equation}%
where $E_{n}$ is given by Eq. (49).

\newpage

{\normalsize %center
}

\bigskip \bigskip

\bigskip \newpage

\bigskip

\ {\normalsize %center
}

\baselineskip= 2\baselineskip% double space the text
%\end{document}
\ 
\begin{table}[tbp]
\caption{ The spectroscopic constants of the EP.for $H_{2}$ and $Ar_{2}$
molecules [7].}%
\begin{tabular}{lll}
\tableline Parameters & $H_{2}$ & $Ar_{2}$ \\ 
\tableline\tableline$D_{e}$ $(cm^{-1})$ & 38281 & 99.55 \\ 
$r_{e}$ $(A^{\circ })$ & 0.7414 & 3.759 \\ 
$\mu $ (a.m.u) & 0.50407 & 19.9812 \\ 
\tableline &  & 
\end{tabular}%
\end{table}
\bigskip \bigskip

\begin{table}[tbp]
\caption{The EP parameters of the $V_{+}(r)$ and the ground state energy, $%
E_{+}^{00}$ (in $cm^{-1}$) of the $H_{2}$ molecule.}%
\begin{tabular}{llllll}
\tableline$\sigma $ & $\delta $ & $\alpha $ $(A^{\circ })^{-1}$ & $E_{+}(SC)$
& $E_{+}(QM)$ & Present \\ 
\tableline\tableline$426.826$ & $463.102$ & $0.9327$ & $2167.68$ & $2168.93$
& 2168.68 \\ 
$47.294$ & $102.341$ & $0.6146$ & $2153.69$ & $2164.83$ & 2164.45 \\ 
$28.685$ & $117.121$ & $0.3826$ & $2139.57$ & $2157.69$ & 2157.53 \\ 
$21.250$ & $213.212$ & $0.1762$ & $2124.29$ & $2148.40$ & 2147.53 \\ 
\tableline &  &  &  &  & 
\end{tabular}%
\end{table}
\begin{table}[tbp]
\caption{Comparisons of experimentally calculated $s$-states energy
transition values $\Delta E_{n,0}(cm^{-1})$ for $n\neq 0\rightarrow n=0$
together with the results of the SC procedure and the present NU method for
the $Ar_{2}$ molecule.}%
\begin{tabular}{lllllll}
\tableline$n$ & Present & $\Delta E(a)$ & $\Delta E(b)$ & $\Delta E(c)$ & $%
\Delta E(d)$ & $\Delta E_{+}(SC)$ \\ 
\tableline\tableline1 & 25.808 & 25.74 & 25.49 & 25.21 & 25.56 & 25.75 \\ 
2 & 46.079 & 46.15 & 45.63 & 45.02 & 46.00 & 46.01 \\ 
3 & 61.472 & 61.75 & 60.70 & 60.04 & 61.32 & 61.42 \\ 
4 & 72.536 & 72.66 & 71.33 & 70.92 & 71.52 & 72.52 \\ 
5 & 79.733 & 79.44 & - & - & - & 79.79 \\ 
6 & 83.453 & - & - & - & - & 83.59 \\ 
7 & 84.026 & - & - & - & - & - \\ 
\tableline &  &  &  &  &  & 
\end{tabular}%
\end{table}

\begin{table}[tbp]
\caption{Energy levels $E_{n,l}(cm^{-1})$ for $Ar_{2}$ and $H_{2}$ molecules
in $V_{+}(r)$ u$\sin $g the NU method.}%
\begin{tabular}{llll}
\tableline$n$ & $l$ & $E_{+}(Ar_{2})$ & $E_{+}(H_{2})$ \\ 
\tableline\tableline0 & 0 & 15.3828 & 2168.68 \\ 
1 & 0 & 41.1910 & 6306.66 \\ 
& 1 & 25.7584 & 6331.10 \\ 
2 & 0 & 61.4619 & 10183.8 \\ 
& 1 & 49.7874 & 10207.6 \\ 
& 2 & - & 10255.2 \\ 
3 & 0 & 76.8546 & 13802.1 \\ 
& 1 & 68.3028 & 13825.2 \\ 
& 2 & 19.9133 & 13871.5 \\ 
4 & 0 & 87.9188 & 17163.2 \\ 
& 1 & 82.0041 & 17185.7 \\ 
& 2 & 46.4777 & 17230.7 \\ 
5 & 0 & 95.1159 & 20269.1 \\ 
& 1 & 91.4672 & 20291.0 \\ 
& 2 & 66.5474 & 20334.8 \\ 
\tableline &  &  & 
\end{tabular}%
\end{table}

\bigskip \bigskip

\end{document}